\pdfoutput=1
\documentclass[10pt,conference]{IEEEtran}
\IEEEoverridecommandlockouts

\usepackage[T1]{fontenc}
\usepackage[utf8]{inputenc}
\usepackage{amsmath,amssymb}
\usepackage{graphicx}
\usepackage{booktabs}
\usepackage{xcolor}
\usepackage{pifont}
\usepackage{listings}
\usepackage{tikz}
\usetikzlibrary{arrows.meta}
\usepackage{capt-of}
\usepackage{cite}
\usepackage[hidelinks]{hyperref}
\usepackage{xurl}
\hypersetup{
  pdftitle={Teralizer: A Tool for Generalizing JUnit Tests into Property-Based jqwik Tests},
  pdfauthor={Johann Glock, Clemens Bauer, Martin Pinzger},
  pdflang={en-US}
}

\newcommand{\Teralizer}{\textsc{Teralizer}}
\newcommand{\codefont}{\fontfamily{lmtt}\selectfont}
\newcommand{\code}[1]{{\codefont #1}}
\newcommand{\runin}[1]{\textit{#1:}}
\newcommand{\Naive}{\textsc{Naive}}
\newcommand{\Improved}{\textsc{Improved}}

\lstdefinestyle{code}{
  basicstyle=\codefont\fontsize{7.6}{9}\selectfont,
  columns=fullflexible,
  keepspaces=true,
  showstringspaces=false,
  aboveskip=4pt,
  belowskip=4pt,
  xleftmargin=1em
}

\lstdefinestyle{console}{
  basicstyle=\codefont\fontsize{5.5}{6.6}\selectfont,
  columns=fullflexible,
  keepspaces=true,
  showstringspaces=false,
  literate={-}{{-}}1,
  frame=single,
  framerule=0.4pt,
  framesep=2pt,
  rulecolor=\color{black!22},
  backgroundcolor=\color{black!2},
  xleftmargin=2.4pt,
  xrightmargin=2.4pt,
  aboveskip=0pt,
  belowskip=2pt
}

\definecolor{inhl}{RGB}{212,227,247}
\definecolor{outhl}{RGB}{253,224,190}
\definecolor{delhl}{RGB}{250,214,214}
\definecolor{addhl}{RGB}{212,239,212}
\definecolor{delmark}{RGB}{240,168,168}
\definecolor{addmark}{RGB}{158,214,158}
\newcommand{\slip}[2]{\hspace{-\fboxsep}\colorbox{#1}{#2}\hspace{-\fboxsep}}
\definecolor{effone}{RGB}{13,122,120}
\definecolor{efften}{RGB}{52,84,160}
\definecolor{effsixty}{RGB}{194,65,12}
\tikzset{
  paperfig/.style={x=1pt, y=1pt, font=\sffamily,
    execute at begin picture={\setlength{\fboxsep}{1.2pt}},
    panel/.style={draw=black!30, line width=0.5pt, rounded corners=2.5pt, fill=white},
    subbox/.style={draw=black!22, line width=0.4pt, rounded corners=1.5pt, fill=black!2},
    num/.style={circle, fill=black, text=white, inner sep=0pt, minimum size=8.2pt,
      font=\sffamily\bfseries\fontsize{5.8}{6}\selectfont},
    ttl/.style={anchor=base west, yshift=-2.5pt, inner sep=0pt, font=\sffamily\bfseries\fontsize{7.1}{8.1}\selectfont},
    code/.style={anchor=west, inner sep=0pt, font=\fontfamily{lmtt}\fontsize{5.7}{6.9}\selectfont},
    txt/.style={anchor=west, inner sep=0pt, font=\sffamily\fontsize{6.0}{7.1}\selectfont},
    tag/.style={anchor=west, inner sep=0pt, font=\sffamily\fontsize{5.7}{6.7}\selectfont},
    hdr/.style={txt, font=\sffamily\bfseries\fontsize{5.7}{6.7}\selectfont, text=black!65},
    spec/.style={anchor=west, inner xsep=2.5pt, inner ysep=1.6pt, rounded corners=1.2pt,
      font=\sffamily\fontsize{6.2}{7.2}\selectfont},
    badge/.style={tag, line width=0.3pt, rounded corners=1.2pt, inner xsep=2.5pt, inner ysep=1.2pt},
    flow/.style={-{Triangle[length=3.0pt, width=4.6pt]}, line width=1.0pt, black!60}
  }
}

\title{Teralizer: A Tool for Generalizing JUnit Tests\\into Property-Based jqwik Tests}

\author{%
\IEEEauthorblockN{Johann Glock}
\IEEEauthorblockA{\textit{Software Competence Center}\\
\textit{Hagenberg (SCCH)}\\
Hagenberg, Austria\\
0000-0002-0152-8611}
\and
\IEEEauthorblockN{Clemens Bauer}
\IEEEauthorblockA{\textit{Department of Informatics Systems}\\
\textit{University of Klagenfurt}\\
Klagenfurt, Austria\\
0009-0000-9199-8563}
\and
\IEEEauthorblockN{Martin Pinzger}
\IEEEauthorblockA{\textit{Department of Informatics Systems}\\
\textit{University of Klagenfurt}\\
Klagenfurt, Austria\\
0000-0002-5536-3859}}

\newsavebox{\ideabox}
\IEEEaftertitletext{%
  \setbox\ideabox=\vbox{%
    \vspace*{-0.4\baselineskip}%
    {\centering\begin{tikzpicture}[paperfig]
\draw[rounded corners=4pt, fill=black!3, draw=black!20, line width=0.4pt] (0,0) rectangle (516,-176.6);
\draw[panel] (5,-5) rectangle (208,-171.6);
\draw[panel] (217,-5) rectangle (511,-66.6);
\draw[panel] (217,-74.6) rectangle (511,-171.6);
\draw[flow] (208.8,-36) -- (216.4,-36);
\draw[flow] (364,-67.4) -- (364,-74);

\node[num] at (14,-14) {1};
\node[ttl] at (21,-14) {Method under test and unit test};

\draw[subbox] (9,-23) rectangle (204,-107.6);
\node[hdr] at (12.5,-29.5) {Method under test};
\node[tag, anchor=east, text=black!65] at (200.5,-29.5) {$-$/$+$: refactoring};
\fill[delhl] (10.2,-80.6) rectangle (202.8,-57.8);
\fill[addhl] (10.2,-95.8) rectangle (202.8,-80.6);
\foreach \y in {-61.6, -69.2, -76.8} \node[code] at (12.5,\y) {-};
\foreach \y in {-84.4, -92.0} \node[code] at (12.5,\y) {+};
\node[code] at (18.6,-38.8) {static final int FULL = 60, PARTIAL = 5;};
\node[code] at (18.6,-46.4) {public static int approvedAmount(int score, int amount) \{};
\node[code] at (18.6,-54.0) {\ \ if (amount <= 0) throw new IllegalArgumentException();};
\node[code] at (18.6,-61.6) {\ \ if (\slip{delmark}{score >= 80}) return Math.min(amount, FULL);};
\node[code] at (18.6,-69.2) {\ \ if (score >= 50) return Math.min(amount, PARTIAL);};
\node[code] at (18.6,-76.8) {\ \ return 0;};
\node[code] at (18.6,-84.4) {\ \ int limit = \slip{addmark}{score > 80} ? FULL : score >= 50 ? PARTIAL : 0;};
\node[code] at (18.6,-92.0) {\ \ return Math.min(amount, limit);};
\node[code] at (18.6,-99.6) {\}};

\draw[subbox] (9,-111.6) rectangle (204,-148.6);
\node[hdr] at (12.5,-118.1) {Unit test};
\node[tag, anchor=east, text=black!65] at (200.5,-118.1) {validates a single input-output pair};
\node[code] at (12.5,-127.0) {@Test void fullApproval() \{};
\node[code] at (12.5,-134.6) {\ \ assertEquals(\colorbox{outhl}{20}, LoanPolicy.approvedAmount(\colorbox{inhl}{90, 20}));};
\node[code] at (12.5,-142.2) {\}};

\node[tag] at (9,-156.1) {Before the refactoring: \textcolor{green!35!black}{\ding{51}\ passes}};
\node[tag] at (9,-164.6) {After the refactoring: \textcolor{green!35!black}{\ding{51}\ passes}, i.e., the \textbf{regression is not detected}};

\node[num] at (226,-14) {2};
\node[ttl] at (233,-14) {Input/output specification extracted by \Teralizer{}};

\node[hdr] at (221,-29) {Path condition};
\node[spec, fill=inhl] at (273,-29) {\textit{amount} > 0 \ $\land$\ \ \textit{score} $\ge$ 80 \ $\land$\ \ \textit{amount} $\le$ 60};
\node[tag, text=black!65] at (410,-29) {inputs on the same execution path};
\node[hdr] at (221,-42.5) {Symbolic output};
\node[spec, fill=outhl] at (273,-42.5) {\textit{amount}};
\node[tag, text=black!65] at (410,-42.5) {expected output for these inputs};
\node[tag, text=black!65] at (221,-57) {Extracted from the original implementation through single-path symbolic analysis of the test execution.};

\node[num] at (226,-84) {3};
\node[ttl] at (233,-84) {Generalized test created by \Teralizer{} (simplified)};

\draw[subbox] (221,-93) rectangle (431,-144.5);
\node[code] at (224.5,-100.6) {@Property};
\node[code] at (224.5,-108.2) {void fullApproval(};
\node[code] at (224.5,-115.8) {\ \ \ \ @ForAll \colorbox{inhl}{@IntRange(min = 80)} int score,};
\node[code] at (224.5,-123.4) {\ \ \ \ @ForAll \colorbox{inhl}{@IntRange(min = 1, max = 60)} int amount) \{};
\node[code] at (224.5,-131.0) {\ \ assertEquals(\colorbox{outhl}{amount}, LoanPolicy.approvedAmount(score, amount));};
\node[code] at (224.5,-138.6) {\}};

\fill[inhl] (437,-114.5) rectangle (442,-109.5);
\node[tag] at (445,-112) {generated inputs satisfy};
\node[tag] at (445,-119.4) {the path condition};
\fill[outhl] (437,-130.5) rectangle (442,-125.5);
\node[tag] at (445,-128) {symbolic output replaces};
\node[tag] at (445,-135.4) {the expected value};

\node[tag] at (221,-156.1) {Before the refactoring: \textcolor{green!35!black}{\ding{51}\ passes}};
\node[tag] at (221,-164.6) {After the refactoring: \textcolor{red!65!black}{\ding{55}\ fails} for \textit{score} = 80, \textit{amount} = 60 (expected 60 but was 5), i.e., the \textbf{regression is detected}};
\end{tikzpicture}
\par}%
    \vspace{0.3\baselineskip}%
    \captionof{figure}{Test generalization with \Teralizer{} on the running example. (1)~The unit test validates a single input and passes before and after a refactoring that unintentionally changes the condition \code{score >= 80} to \code{score > 80}. (2)~\Teralizer{} extracts the path condition and the symbolic output expression of the tested execution path. (3)~The generalized test validates the assertion for generated inputs that satisfy the path condition and thereby detects the regression.}%
    \label{fig:idea}}%
  \nointerlineskip\box\ideabox\vspace*{\baselineskip}}

\begin{document}

\maketitle

\begin{abstract}
\textbf{Background:} Property-based testing validates properties across many generated inputs and can thereby detect regressions that conventional unit tests miss.
However, writing property-based tests is challenging for practitioners, which limits their adoption.
\textbf{Tool:} \Teralizer{} automatically generalizes JUnit tests into property-based jqwik tests for regression detection.
Through single-path symbolic analysis, it extracts from each test execution a path condition, which characterizes the inputs that follow the same execution path, and a symbolic output expression, which computes the expected output for these inputs.
The generalized tests generate inputs that satisfy the path condition and use the symbolic output expression as the oracle.
\textbf{Results:} In our evaluation, generalization increased the mutation scores of EvoSuite-generated test suites by 0.8 to 3.9 percentage points.
On the EqBench benchmark, generalizing the test suites that EvoSuite generated with 1\,s per class reached the mutation score of EvoSuite's 60\,s test suites with 31.9\% less total runtime.
\textbf{Availability:} \Teralizer{} is available as open source at \url{https://github.com/glockyco/Teralizer}.
\end{abstract}


\begin{figure*}[t]
\centering
\begin{tikzpicture}[paperfig,
  sname/.style={anchor=west, inner sep=0pt, align=left, font=\sffamily\bfseries\fontsize{6.3}{7.2}\selectfont},
  tool/.style={tag, text=black!60},
  colhdr/.style={anchor=west, inner sep=0pt, font=\sffamily\fontsize{5.4}{6.3}\selectfont, text=black!60},
  cell/.style={anchor=west, inner sep=0pt, align=left, font=\sffamily\fontsize{5.7}{7.0}\selectfont},
  pc/.style={anchor=west, align=left, inner xsep=2.2pt, inner ysep=1.5pt, rounded corners=1.2pt, fill=inhl,
    font=\sffamily\fontsize{5.7}{6.8}\selectfont},
  outval/.style={anchor=west, inner xsep=2.2pt, inner ysep=1.3pt, rounded corners=1.2pt,
    font=\sffamily\fontsize{5.7}{6.8}\selectfont},
  rowcode/.style={anchor=west, inner sep=0pt, font=\fontfamily{lmtt}\fontsize{5.6}{6.7}\selectfont},
  nextstage/.style={-{Triangle[length=2.6pt, width=3.8pt]}, line width=0.8pt, black!60},
  guide/.style={black!13, line width=0.3pt},
]
\draw[rounded corners=4pt, fill=black!3, draw=black!20, line width=0.4pt] (0,0) rectangle (516,-130.5);

\draw[panel, fill=black!6] (5,-5) rectangle (61,-30);
\node[sname] at (9.5,-13.2) {Java project};
\node[tool] at (9.5,-22.6) {with JUnit tests};
\foreach \x/\w/\n/\name/\uses in {
    67/80/1/{Test and Assertion\\Analysis}/Spoon,
    153/80/2/{Tested Method\\Identification}/Spoon,
    239/108/3/{Specification\\Extraction}/SPF,
    353/86/4/{Generalized Test\\Creation}/jqwik,
    445/66/5/{Test Suite\\Reduction}/{PIT, optional}} {
  \draw[panel] (\x,-5) rectangle (\x+\w,-30);
  \node[num] at (\x+8.5,-10.9) {\n};
  \node[sname] at (\x+15.5,-14.3) {\name};
  \node[tool] at (\x+15.5,-26) {\uses};
}
\foreach \x in {61.7, 147.7, 233.7, 347.7, 439.7} \draw[nextstage] (\x,-17.5) -- (\x+4.8,-17.5);

\draw[panel] (5,-34) rectangle (511,-125.5);
\draw[guide] (5,-44.5) -- (511,-44.5);
\foreach \y in {-71.5, -98.5} \draw[guide] (5,\y) -- (511,\y);
\foreach \x in {64, 236, 350, 442} \draw[guide] (\x,-34) -- (\x,-125.5);
\node[colhdr] at (8.5,-39.5) {Unit test};
\node[colhdr] at (70.5,-39.5) {Assertion with MUT call (underlined)};
\node[colhdr] at (242.5,-39.5) {Path condition and output from SPF};
\node[colhdr] at (356.5,-39.5) {Oracle for the input partition};
\node[colhdr] at (448.5,-39.5) {Additional mutants};

\node[rowcode] at (8.5,-58) {fullApproval};
\node[rowcode] at (70.5,-58) {assertEquals(\slip{outhl}{20}, \underline{LoanPolicy.approvedAmount(\slip{inhl}{90, 20})});};
\node[pc] at (240.5,-53.7) {\textit{amount} > 0 \,$\land$\, \textit{score} $\ge$ 80\\$\land$\, \textit{amount} $\le$ 60};
\node[cell] at (242.7,-66.9) {output:};
\node[outval, fill=outhl] at (262.5,-66.9) {\textit{amount}};
\node[cell, text=black!60] at (288,-66.9) {(symbolic)};
\node[cell] at (356.5,-58) {\textcolor{green!35!black}{\ding{51}}\ symbolic output\\$\rightarrow$ generalized test (Fig.~\ref{fig:idea})};
\node[cell] at (448.5,-58) {\textcolor{green!35!black}{\ding{51}}\ 1 additional mutant\\$\rightarrow$ retained};

\node[rowcode] at (8.5,-85) {partialApproval};
\node[rowcode] at (70.5,-85) {assertEquals(\slip{outhl}{5}, \underline{LoanPolicy.approvedAmount(\slip{inhl}{60, 8})});};
\node[pc] at (240.5,-80.7) {\textit{amount} > 0 \,$\land$\, \textit{score} < 80\\$\land$\, \textit{score} $\ge$ 50 \,$\land$\, \textit{amount} > 5};
\node[cell] at (242.7,-93.9) {output:};
\node[outval, fill=black!9] at (262.5,-93.9) {5};
\node[cell, text=black!60] at (273,-93.9) {(concrete)};
\node[cell] at (356.5,-85) {\textcolor{red!65!black}{\ding{55}}\ output 5 is returned via\\\code{Math.min}, not as a literal\\$\rightarrow$ excluded};
\node[cell, text=black!60] at (448.5,-85) {--};

\node[rowcode] at (8.5,-112) {rejection};
\node[rowcode] at (70.5,-112) {assertEquals(\slip{outhl}{0}, \underline{LoanPolicy.approvedAmount(\slip{inhl}{30, 1})});};
\node[pc] at (240.5,-107.7) {\textit{amount} > 0 \,$\land$\, \textit{score} < 80\\$\land$\, \textit{score} < 50};
\node[cell] at (242.7,-120.9) {output:};
\node[outval, fill=black!9] at (262.5,-120.9) {0};
\node[cell, text=black!60] at (273,-120.9) {(concrete)};
\node[cell] at (356.5,-112) {\textcolor{green!35!black}{\ding{51}}\ output 0 is returned\\as a literal\\$\rightarrow$ generalized test};
\node[cell] at (448.5,-112) {\textcolor{red!65!black}{\ding{55}}\ no additional mutant\\$\rightarrow$ not retained};
\end{tikzpicture}
\caption{The five stages of \Teralizer{} with the tools that they use, and their results for the three unit tests of the running example.}
\label{fig:pipeline}
\end{figure*}

\section{Introduction}
\label{sec:introduction}

Unit tests are widely used to detect regressions, i.e., unintended changes in program behavior.
Since a conventional unit test validates the method under test (MUT) only for specific input-output pairs, regressions that affect other inputs remain undetected.
Fig.~\ref{fig:idea} shows an example.
The method \code{approvedAmount} approves loans of up to \code{FULL} for applicants with a credit score of at least 80, and the unit test \code{fullApproval} validates this behavior for a score of 90 and a requested amount of 20.
A refactoring that merges the two calls of \code{Math.min} unintentionally changes the condition \code{score >= 80} to \code{score > 80}.
Applicants with a score of 80 are thereby approved for at most \code{PARTIAL} instead of \code{FULL}, but the unit test \code{fullApproval} still passes.

Property-based testing (PBT)~\cite{claessen_2000_quickcheck} addresses this limitation by validating properties across many inputs that are produced by generators.
In the example, a property-based test that validates the approved amount for all scores of at least 80 detects the regression.
However, writing property-based tests requires developers to identify suitable properties and to define generators with appropriate constraints, which practitioners find challenging and which limits the adoption of PBT~\cite{goldstein_2024_pbt_practice}.

In this paper, we present \Teralizer{}, a tool that automatically generalizes JUnit tests into property-based jqwik~\cite{link_2022_jqwik} tests based on the approach of our previous work~\cite{glock_2025_teralizer}.
Through single-path symbolic analysis with Symbolic PathFinder (SPF)~\cite{pasareanu_2013_symbolic}, \Teralizer{} extracts from the execution of a unit test a \emph{path condition}, which characterizes all inputs that follow the same execution path as the original inputs, and a \emph{symbolic output expression}, which computes the expected output for these inputs.
It then creates a property-based test that generates inputs satisfying the path condition and uses the symbolic output expression as the expected value of the original assertion.
In Fig.~\ref{fig:idea}, this generalized test detects the regression that the unit test misses.

The approach and its evaluation are described in detail in~\cite{glock_2025_teralizer}, while this paper focuses on the implementation and the usage of the tool.
\Teralizer{} is open source under the MIT license,\footnote{\url{https://github.com/glockyco/Teralizer}} with a replication package on Zenodo~\cite{glock_2026_replication}.

\section{Overview of \Teralizer{}}
\label{sec:approach}

\Teralizer{} is implemented in Java and takes a Maven or Gradle project with JUnit~3, 4, or~5 tests as input, to which it adds the generalized tests as jqwik test classes.
Fig.~\ref{fig:pipeline} shows the five stages of its fully automated pipeline~\cite{glock_2025_teralizer} and their results for the three unit tests of the running example.
Tests and assertions that a stage cannot process are excluded from the subsequent stages, and the reasons are recorded.

\runin{Stages 1 and 2}
\Teralizer{} executes the test suite and analyzes all passing tests with Spoon~\cite{pawlak_2016_spoon}.
Stage~1 identifies the assertions that \Teralizer{} can generalize, which currently are equality, Boolean, and exception assertions of JUnit and equality assertions of Hamcrest.
Stage~2 identifies the MUT call of each assertion, i.e., the method call whose result the assertion validates.
For this purpose, a data-flow analysis traces the asserted value back to the method call that produced it.
If the analysis does not yield a unique call, a ranking heuristic selects one of the preceding calls based on its parameter and return types, its class, its name, and its distance to the assertion.
Arguments of primitive types, their boxed types, and strings become the inputs of the generalized test.

\runin{Stage 3}
\Teralizer{} executes each test with SPF using single-path symbolic analysis, which treats these arguments as symbolic values while it follows the execution path of the original inputs and therefore requires no constraint solving.
SPF adds the condition of each taken branch that depends on a symbolic value to the path condition.
When the MUT returns, \Teralizer{} records the path condition and the output as the input/output specification.
For \code{fullApproval}, the output is the symbolic expression $\mathit{amount}$, whereas SPF provides only the concrete outputs 5 and~0 for the other two tests.

\runin{Stage 4}
\Teralizer{} creates a jqwik test class with a property-based version of the original test (Fig.~\ref{fig:idea}), in which the arguments of the MUT call are replaced with generated parameters and the expected value of the assertion with the output of the specification.
The output therefore has to be a valid oracle for all inputs of the input partition, which holds for a symbolic output expression.
A concrete output is not necessarily constant for the input partition, because SPF also provides only a concrete output when it cannot track the dependence of the output on the inputs, e.g., for outputs that are accumulated in input-dependent loops~\cite{glock_2025_teralizer}.
\Teralizer{} therefore uses a concrete output only if the MUT directly returns a literal, as in \code{rejection} (\code{return 0}).
In \code{partialApproval}, the constant is returned via \code{Math.min}, which the current implementation does not recognize as constant, so that the assertion is excluded.

The parameters are generated according to one of two strategies.
\Naive{} generates random values and applies the path condition as a filter.
\Improved{} additionally encodes constraints that compare a parameter with a constant or with another parameter in the generators, which reduces the number of rejected inputs and enables jqwik to generate inputs at the boundaries of the input partition, such as the score of 80 that reveals the regression in Fig.~\ref{fig:idea}.
The figure shows these constraints as jqwik annotations, whereas the generated code implements them in a supplier class.
A generalized test first executes the original inputs and then generated inputs, up to a configurable number of inputs per test (tries).
Generalized tests that fail for the current implementation are excluded.

\runin{Stage 5}
This optional stage applies mutation testing with PIT~\cite{coles_2016_pit} and retains a generalized test only if it detects at least one mutant that the original test suite does not detect.
In the example, this holds for \code{fullApproval}, whose generalized test detects the mutant that changes \mbox{\code{score >= 80}} to \mbox{\code{score > 80}}, i.e., the fault that the refactoring of Fig.~\ref{fig:idea} introduces.

\begin{figure}[t]
\begin{lstlisting}[style=console]
==> Outcome of each assertion that reached generalization
      test       | tested_method  | generalized |              reason
-----------------+----------------+-------------+----------------------------------
 fullApproval    | approvedAmount | t           |
 partialApproval | approvedAmount | f           | NULL_CONCRETE_OUTPUT_NOT_LITERAL
 rejection       | approvedAmount | t           |
[...]
==> Running all tests against the refactored example
[INFO] Tests run: 3, Failures: 0, Errors: 0, Skipped: 0 -- in bank.LoanPolicyTest
    expected: <60> but was: <5>
tries = 4                     | # of calls to property
  arg0: TestParameters{score=80,amount=60}
[ERROR] Tests run: 1, Failures: 1, [...] -- in [...]_Generalized_fullApproval_2_Test
[...]
\end{lstlisting}
\caption{Output of \code{scripts/try-example.sh} for the example (abridged).}
\label{fig:output}
\end{figure}

\section{Usage}
\label{sec:usage}

\Teralizer{} is executed in a Docker container.
On a machine with Bash, Git LFS, and Docker, the following commands build the image in a clone of the repository and apply \Teralizer{} to the running example:
\begin{lstlisting}[style=code]
git lfs pull
git submodule update --init --recursive jpf-symbc
docker build -t teralizer .
scripts/try-example.sh
\end{lstlisting}
\noindent The script \code{try-example.sh} runs the pipeline on a copy of the example and prints the results (Fig.~\ref{fig:output}).
It reports for each assertion that reaches Stage~4 whether a generalized test was created or why the assertion was excluded, the mutants that the original and the generalized test suite detect, and the paths of the generalized test classes.
It then applies the refactoring of Fig.~\ref{fig:idea} and executes all tests.
The three original tests pass, whereas the generalized test of \code{fullApproval} fails at the fourth try and reports the failing input together with the expected and the actual output.
Processing the example takes less than two minutes on a laptop (Intel Core Ultra~7 265H, 16\,GB of memory assigned to Docker).

For another project, the script takes the project directory as its argument.
The project has to build with Java~8, which SPF requires, and use Maven or Gradle with JUnit~3, 4, or~5.
The script works on a copy of the project and uses the configuration file of the running example, which specifies the generalization strategy and the number of tries per generalized test:
\begin{lstlisting}[style=code]
teralizer.generalizations.IMPROVED_200_TRIES {
  algorithm = "IMPROVED", jqwik.tries = 200
}
\end{lstlisting}

\Teralizer{} adds each generalized test as a separate jqwik test class to the package of the original test class and does not modify existing tests.
The generalized tests are executed on the JUnit Platform.
In the running example, adding jqwik as a test dependency is sufficient to execute them together with the original tests via \code{mvn test}.
The results of all stages and the reasons for all exclusions are stored in a PostgreSQL database.

\begin{figure}[t]
\centering
\begin{tikzpicture}[x=1pt, y=1pt, font=\sffamily\fontsize{6.2}{7.2}\selectfont,
  grid/.style={black!12, line width=0.3pt},
  axis/.style={black!55, line width=0.4pt},
  tick/.style={inner sep=0pt, text=black!75},
  name/.style={inner sep=0pt, anchor=north west, font=\sffamily\bfseries\fontsize{6.6}{7.6}\selectfont},
  chain/.style={line width=0.9pt, line join=round},
  b1/.style={color=effone}, b10/.style={color=efften}, b60/.style={color=effsixty},
  alone/.style={fill=white, line width=0.9pt},
  gain/.style={black, line width=0.5pt, -{Triangle[length=2.6pt, width=3pt]}, shorten <=3.4pt, shorten >=3pt},
  note/.style={inner sep=0pt, font=\sffamily\bfseries\fontsize{6.2}{7}\selectfont},
  dot/.pic={\fill (0,0) circle[radius=1.7pt];},
  sq/.pic={\fill (-1.5pt,-1.5pt) rectangle (1.5pt,1.5pt);},
  tri/.pic={\fill (-1.9pt,-1.3pt) -- (1.9pt,-1.3pt) -- (0,2pt) -- cycle;},
]
\useasboundingbox (0,-5) rectangle (252,111);
\node[tick, anchor=west] at (0,106) {EvoSuite search budget per class:};
\draw[b1, alone] (104,106) circle[radius=2.1pt];
\node[tick, anchor=west] at (108.5,106) {1\,s};
\draw[b10, alone] (123,103.9) rectangle (127.2,108.1);
\node[tick, anchor=west] at (130,106) {10\,s};
\draw[b60, alone] (148,104) -- (152.8,104) -- (150.4,108.4) -- cycle;
\node[tick, anchor=west] at (155.5,106) {60\,s};
\node[tick, anchor=west] at (0,97) {Outlined markers: EvoSuite alone. Filled markers: EvoSuite followed by \Teralizer{}.};

\node[tick, rotate=90] at (3,49) {mutation score};

\begin{scope}[shift={(-12,-400.2)}, x=1.6667pt, y=8.7805pt]
  \foreach \det in {48, 50, 52, 54} {
    \draw[grid] (21.6,\det) -- (97.2,\det);
    \node[tick, anchor=east] at (20.1,\det) {\det\%};}
  \draw[axis] (21.6,47.4) -- (97.2,47.4);
  \foreach \hours/\ks in {10/36, 15/54, 20/72, 25/90}
    \node[tick, anchor=north] at (\ks,46.99) {\hours\,h};
  \node[tick, anchor=north] at (59.4,46.0) {total runtime};
  \node[name] at (23,55.55) {EqBench};

  \draw[chain, b1] (26.479,48.1) -- (36.728,50.7) -- (37.457,51.4) -- (37.532,51.7);
  \draw[chain, b10] (29.861,50.6) -- (41.525,51.9) -- (42.256,53.6) -- (45.398,53.8) -- (48.269,53.8) -- (62.938,54.1);
  \draw[chain, b60] (55.075,51.6) -- (68.093,54.5) -- (68.782,54.7) -- (75.081,54.8) -- (93.017,55.0);
  \foreach \point in {(36.728,50.7), (37.457,51.4), (37.532,51.7)} \pic[b1] at \point {dot};
  \foreach \point in {(41.525,51.9), (42.256,53.6), (45.398,53.8), (48.269,53.8), (62.938,54.1)} \pic[b10] at \point {sq};
  \foreach \point in {(68.093,54.5), (68.782,54.7), (75.081,54.8), (93.017,55.0)} \pic[b60] at \point {tri};
  \draw[b1, alone] (26.479,48.1) circle[radius=2.1pt];
  \draw[b10, alone] (29.861,50.6) +(-2.1pt,-2.1pt) rectangle +(2.1pt,2.1pt);
  \draw[b60, alone] (55.075,51.6) +(-2.4pt,-1.7pt) -- +(2.4pt,-1.7pt) -- +(0,2.6pt) -- cycle;

  \draw[gain] (55.075,51.6) -- (37.532,51.7);
  \draw[gain] (55.075,51.6) -- (48.269,53.8);
  \node[note, anchor=south east] at ([xshift=-2.6pt, yshift=1pt]37.532,51.7) {A};
  \node[note, anchor=south] at ([yshift=2.8pt]48.269,53.8) {B};
  \node[tick, anchor=south east, align=left, fill=white, inner sep=1.5pt] at (96.9,47.57)
    {Compared to EvoSuite with 60\,s:\\\textbf{A}: +0.1 pp in 31.9\% less time\\\textbf{B}: +2.2 pp in 12.4\% less time};
\end{scope}

\begin{scope}[shift={(150.92,-1216.73)}, x=5.8559pt, y=21.818pt]
  \foreach \det in {57, 58, 59} {
    \draw[grid] (3.6,\det) -- (16.92,\det);
    \node[tick, anchor=east] at (3.27,\det) {\det\%};}
  \draw[axis] (3.6,56.5) -- (16.92,56.5);
  \foreach \hours/\ks in {2/7.2, 3/10.8, 4/14.4}
    \node[tick, anchor=north] at (\ks,56.335) {\hours\,h};
  \node[tick, anchor=north] at (10.26,55.936) {total runtime};
  \node[name] at (3.85,59.777) {Apache Commons};

  \draw[chain, b1] (4.649,56.8) -- (7.294,57.9);
  \draw[chain, b10] (5.597,57.3) -- (10.445,58.1) -- (10.603,58.4) -- (11.082,58.4) -- (13.270,58.5);
  \draw[chain, b60] (10.240,58.1) -- (13.939,59.3) -- (14.728,59.4) -- (15.736,59.5);
  \pic[b1] at (7.294,57.9) {dot};
  \foreach \point in {(10.445,58.1), (10.603,58.4), (11.082,58.4), (13.270,58.5)} \pic[b10] at \point {sq};
  \foreach \point in {(13.939,59.3), (14.728,59.4), (15.736,59.5)} \pic[b60] at \point {tri};
  \draw[b1, alone] (4.649,56.8) circle[radius=2.1pt];
  \draw[b10, alone] (5.597,57.3) +(-2.1pt,-2.1pt) rectangle +(2.1pt,2.1pt);
  \draw[b60, alone] (10.240,58.1) +(-2.4pt,-1.7pt) -- +(2.4pt,-1.7pt) -- +(0,2.6pt) -- cycle;
\end{scope}
\end{tikzpicture}
\caption{Mutation scores and total runtimes of EvoSuite alone and followed by \Teralizer{}~\cite{glock_2025_teralizer}. Lines connect each EvoSuite test suite with the Pareto-optimal test suites that \Teralizer{} creates from it (pp: percentage points).}
\label{fig:efficiency}
\end{figure}
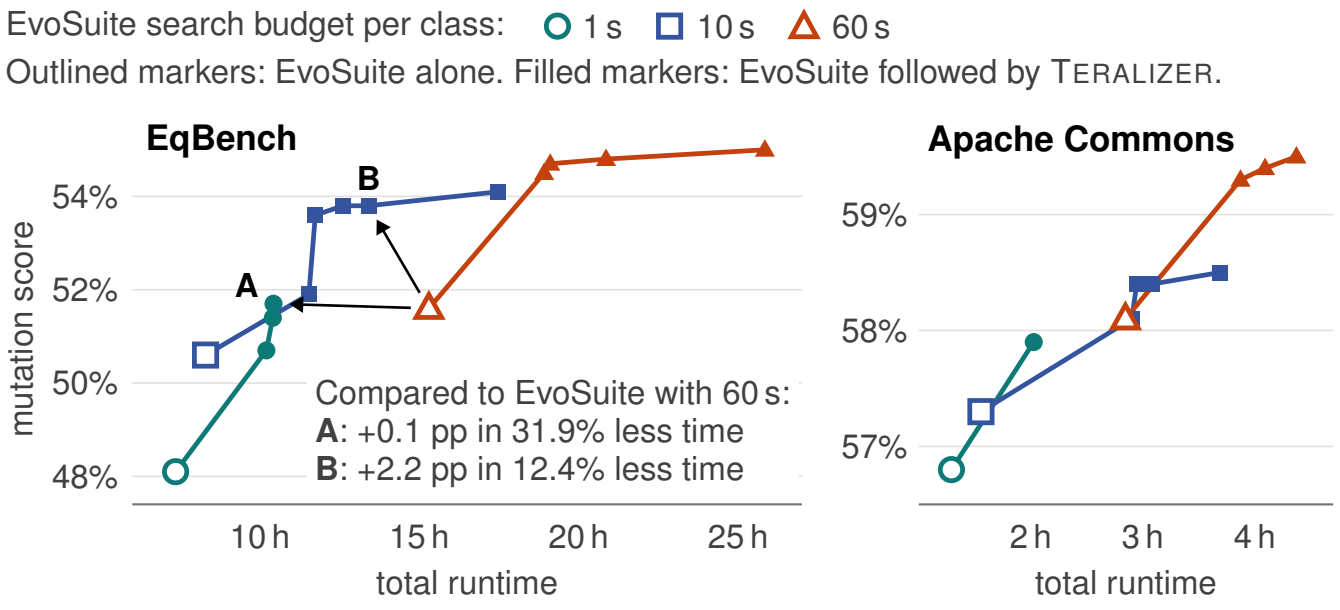

\section{Evaluation}
\label{sec:evaluation}

The evaluation in~\cite{glock_2025_teralizer} uses test suites that EvoSuite~\cite{fraser_2011_evosuite} generates with search budgets of 1\,s, 10\,s, and 60\,s per class for the EqBench benchmark~\cite{badihi_2021_eqbench} and for numeric utility classes of Apache Commons~\cite{apache_commons}, the developer-written tests of the same classes, and 584 open-source projects from the RepoReapers dataset~\cite{munaiah_2017_reporeapers}.
Mutation scores are measured with PIT.

\runin{Effectiveness}
Generalization increases the mutation scores of the EvoSuite-generated test suites by 1.22--3.93 percentage points for EqBench and by 0.82--1.33 percentage points for Apache Commons, depending on the search budget, the generalization strategy, and the number of tries.
For the developer-written tests, which already achieve a mutation score of 80.35\%, the increase is 0.05--0.07 percentage points.
The detection rate increases most for mutants of arithmetic operations and, with \Improved{}, for mutants of conditional boundaries, such as the mutant of the running example.

\runin{Efficiency}
For EqBench, EvoSuite with a search budget of 1\,s followed by \Teralizer{} (A~in Fig.~\ref{fig:efficiency}) achieves a 0.1~percentage points higher mutation score than EvoSuite with a search budget of 60\,s in 31.9\% less total runtime.
With a search budget of 10\,s (B), this combination achieves a 2.2~percentage points higher mutation score in 12.4\% less time.
For Apache Commons, all test suites of EvoSuite are Pareto-optimal, but generalization achieves higher mutation scores than EvoSuite alone (59.5\% vs.\ 58.1\%).
Stage~5 takes 59.1--95.7\% of the processing time of \Teralizer{} and retains 1,555 of the 21,478 generalized tests (7.2\%).

\runin{Applicability}
The current prototype of \Teralizer{} is less successful for real-world projects.
Of the 584 RepoReapers projects, 85 (14.6\%) complete all five stages, which yields 1,435 generalized tests.
Across all projects, 3.8\% of the assertions pass the first three stages, compared with 47.8\% for EqBench and Apache Commons.
The other assertions are excluded mainly because no MUT call can be identified or because the parameters or the return value of the MUT have unsupported types.
In Stage~4, 61.6\% of the generalization attempts are excluded for lack of a valid oracle, 77.4\% of them because of a concrete output that the MUT does not return as a literal, as in \code{partialApproval}.

\section{Limitations and Future Work}
\label{sec:outlook}

Since the oracles of the generalized tests are derived from the implementation, generalized tests detect behavioral changes in later versions, but not faults that are already present in the analyzed version.
Moreover, single-path symbolic analysis requires pure MUTs, i.e., deterministic MUTs without side effects that depend only on their parameters~\cite{glock_2025_teralizer}.

The applicability of \Teralizer{} is mainly limited by the underlying symbolic analysis.
SPF requires Java~8.
Since it constructs exact symbolic models for constraint solving~\cite{pasareanu_2013_symbolic}, it supports only inputs and outputs of primitive types, their boxed types, and strings.
Test generalization has weaker requirements than constraint solving, because a generator that covers only a subset of an input partition reduces the number of tested inputs, but does not cause spurious test failures~\cite{glock_2025_teralizer}.
We therefore plan to extend the symbolic analysis beyond what is required for constraint solving, e.g., with approximate models for arrays, objects, and further string operations.

So far, we have not focused on the presentation of the generalized tests.
For example, the generated test class of \code{fullApproval} comprises 412 lines of code, most of which implement the generators and record the tested inputs for the evaluation, compared to six lines of the simplified version in Fig.~\ref{fig:idea}.
In future work, we plan to make the generalized tests easier to understand and to maintain for practitioners.

\section{Related Work}
\label{sec:related-work}

Test generalization is a form of test amplification~\cite{danglot_2019_snowballing}, which enhances existing tests.
Test generation tools such as EvoSuite~\cite{fraser_2011_evosuite} instead infer test scenarios and assertions from the observed behavior of the implementation, which risks overfitting the implementation~\cite{barr_2015_oracle}.
The setup code and assertions of developer-written tests encode knowledge about the intended behavior that the implementation itself does not contain, which makes their generalized tests less likely to overfit the implementation.
Both techniques can be combined by generalizing generated tests, as in Section~\ref{sec:evaluation}.

To our knowledge, JARVIS~\cite{peleg_2018_jarvis} is the only other tool that automatically generalizes unit tests into property-based tests.
It infers properties from groups of similar tests based on predefined abstraction templates and does not analyze the implementation.
\Teralizer{} creates generalized tests in 9 of the 12 projects of the JARVIS evaluation, whereas the published JARVIS results cover 2 projects~\cite{glock_2025_teralizer}.
Approaches that generate parameterized unit tests start from tests without assertions~\cite{fraser_2011_generating_put} or use values observed at runtime~\cite{tiwari_2024_proze} and do not generalize oracles beyond these values.
Large language models generate property-based tests without existing tests, but their generators and assertions can be invalid or unsound~\cite{vikram_2024_pbt}.

\section{Conclusion}
\label{sec:conclusion}

We presented \Teralizer{}, a tool that automatically generalizes JUnit tests into property-based jqwik tests based on single-path symbolic analysis.
The generalized tests validate the original assertions across the tested input partitions and thereby detect regressions that the original tests miss.

\section*{Acknowledgment}
This research was funded in whole or in part by the Austrian Science Fund (FWF) 10.55776/P36698.
For open access purposes, the author has applied a CC BY public copyright license to any author accepted manuscript version arising from this submission.
The research reported in this paper has been partly funded by BMIMI, BMWET, and the State of Upper Austria in the frame of the SCCH competence center INTEGRATE (FFG grant no.~892418) part of the FFG COMET Competence Centers for Excellent Technologies Programme.
We used OpenAI GPT, Anthropic Claude, and Google Gemini models to assist with writing, coding, and literature search.
The authors reviewed and refined the generated material and take full responsibility for the final manuscript.

\bibliographystyle{IEEEtran}
\bibliography{main}

\begin{thebibliography}{10}
\providecommand{\url}[1]{#1}
\csname url@samestyle\endcsname
\providecommand{\newblock}{\relax}
\providecommand{\bibinfo}[2]{#2}
\providecommand{\BIBentrySTDinterwordspacing}{\spaceskip=0pt\relax}
\providecommand{\BIBentryALTinterwordstretchfactor}{4}
\providecommand{\BIBentryALTinterwordspacing}{\spaceskip=\fontdimen2\font plus
\BIBentryALTinterwordstretchfactor\fontdimen3\font minus
  \fontdimen4\font\relax}
\providecommand{\BIBforeignlanguage}[2]{{%
\expandafter\ifx\csname l@#1\endcsname\relax
\typeout{** WARNING: IEEEtran.bst: No hyphenation pattern has been}%
\typeout{** loaded for the language `#1'. Using the pattern for}%
\typeout{** the default language instead.}%
\else
\language=\csname l@#1\endcsname
\fi
#2}}
\providecommand{\BIBdecl}{\relax}
\BIBdecl

\bibitem{claessen_2000_quickcheck}
K.~Claessen and J.~Hughes, ``{QuickCheck}: A lightweight tool for random
  testing of {Haskell} programs,'' in \emph{Proc. 5th ACM SIGPLAN Int. Conf.
  Functional Programming (ICFP)}.\hskip 1em plus 0.5em minus 0.4em\relax ACM,
  2000, pp. 268--279.

\bibitem{goldstein_2024_pbt_practice}
H.~Goldstein, J.~W. Cutler, D.~Dickstein, B.~C. Pierce, and A.~Head,
  ``Property-based testing in practice,'' in \emph{Proc. 46th IEEE/ACM Int.
  Conf. Softw. Eng. (ICSE)}.\hskip 1em plus 0.5em minus 0.4em\relax ACM, 2024,
  pp. 187:1--187:13.

\bibitem{link_2022_jqwik}
\BIBentryALTinterwordspacing
{jqwik-team}, ``jqwik: Property-based testing in {Java}.'' [Online]. Available:
  \url{https://jqwik.net/}
\BIBentrySTDinterwordspacing

\bibitem{glock_2025_teralizer}
J.~Glock, C.~Bauer, and M.~Pinzger, ``{Teralizer}: Semantics-based test
  generalization from conventional unit tests to property-based tests,'' arXiv
  preprint arXiv:2512.14475, 2025.

\bibitem{pasareanu_2013_symbolic}
C.~S. P{\u{a}}s{\u{a}}reanu, W.~Visser, D.~H. Bushnell, J.~Geldenhuys, P.~C.
  Mehlitz, and N.~Rungta, ``{Symbolic PathFinder}: Integrating symbolic
  execution with model checking for {Java} bytecode analysis,'' \emph{Automated
  Software Engineering}, vol.~20, no.~3, pp. 391--425, 2013.

\bibitem{glock_2026_replication}
\BIBentryALTinterwordspacing
J.~Glock, C.~Bauer, and M.~Pinzger, ``Replication package for: `{Teralizer}:
  Semantics-based test generalization from conventional unit tests to
  property-based tests','' Zenodo, 2026. [Online]. Available:
  \url{https://doi.org/10.5281/zenodo.17950380}
\BIBentrySTDinterwordspacing

\bibitem{pawlak_2016_spoon}
R.~Pawlak, M.~Monperrus, N.~Petitprez, C.~Noguera, and L.~Seinturier,
  ``{Spoon}: A library for implementing analyses and transformations of {Java}
  source code,'' \emph{Software: Practice and Experience}, vol.~46, no.~9, pp.
  1155--1179, 2016.

\bibitem{coles_2016_pit}
H.~Coles, T.~Laurent, C.~Henard, M.~Papadakis, and A.~Ventresque, ``{PIT}: A
  practical mutation testing tool for {Java} (demo),'' in \emph{Proc. 25th ACM
  SIGSOFT Int. Symp. Softw. Testing and Analysis (ISSTA)}.\hskip 1em plus 0.5em
  minus 0.4em\relax ACM, 2016, pp. 449--452.

\bibitem{fraser_2011_evosuite}
G.~Fraser and A.~Arcuri, ``{EvoSuite}: Automatic test suite generation for
  object-oriented software,'' in \emph{Proc. 19th ACM SIGSOFT Symp. Foundations
  of Softw. Eng. and 13th European Softw. Eng. Conf. (ESEC/FSE)}.\hskip 1em
  plus 0.5em minus 0.4em\relax ACM, 2011, pp. 416--419.

\bibitem{badihi_2021_eqbench}
S.~Badihi, Y.~Li, and J.~Rubin, ``{EqBench}: A dataset of equivalent and
  non-equivalent program pairs,'' in \emph{Proc. 18th IEEE/ACM Int. Conf.
  Mining Softw. Repositories (MSR)}.\hskip 1em plus 0.5em minus 0.4em\relax
  IEEE, 2021, pp. 610--614.

\bibitem{apache_commons}
\BIBentryALTinterwordspacing
{The Apache Software Foundation}, ``{Apache Commons}.'' [Online]. Available:
  \url{https://commons.apache.org/}
\BIBentrySTDinterwordspacing

\bibitem{munaiah_2017_reporeapers}
N.~Munaiah, S.~Kroh, C.~Cabrey, and M.~Nagappan, ``Curating {GitHub} for
  engineered software projects,'' \emph{Empirical Software Engineering},
  vol.~22, no.~6, pp. 3219--3253, 2017.

\bibitem{danglot_2019_snowballing}
B.~Danglot, O.~Vera-Perez, Z.~Yu, A.~Zaidman, M.~Monperrus, and B.~Baudry, ``A
  snowballing literature study on test amplification,'' \emph{Journal of
  Systems and Software}, vol. 157, p. 110398, 2019.

\bibitem{barr_2015_oracle}
E.~T. Barr, M.~Harman, P.~McMinn, M.~Shahbaz, and S.~Yoo, ``The oracle problem
  in software testing: A survey,'' \emph{IEEE Transactions on Software
  Engineering}, vol.~41, no.~5, pp. 507--525, 2015.

\bibitem{peleg_2018_jarvis}
H.~Peleg, D.~Rasin, and E.~Yahav, ``Generating tests by example,'' in
  \emph{Proc. 19th Int. Conf. Verification, Model Checking, and Abstract
  Interpretation (VMCAI)}.\hskip 1em plus 0.5em minus 0.4em\relax Springer,
  2018, pp. 406--429.

\bibitem{fraser_2011_generating_put}
G.~Fraser and A.~Zeller, ``Generating parameterized unit tests,'' in
  \emph{Proc. 20th ACM SIGSOFT Int. Symp. Softw. Testing and Analysis
  (ISSTA)}.\hskip 1em plus 0.5em minus 0.4em\relax ACM, 2011, pp. 364--374.

\bibitem{tiwari_2024_proze}
D.~Tiwari, Y.~Gamage, M.~Monperrus, and B.~Baudry, ``{PROZE}: Generating
  parameterized unit tests informed by runtime data,'' in \emph{Proc. 24th IEEE
  Int. Conf. Source Code Analysis and Manipulation (SCAM)}.\hskip 1em plus
  0.5em minus 0.4em\relax IEEE, 2024, pp. 166--176.

\bibitem{vikram_2024_pbt}
V.~Vikram, C.~Lemieux, J.~Sunshine, and R.~Padhye, ``Can large language models
  write good property-based tests?'' arXiv preprint arXiv:2307.04346, 2024.

\end{thebibliography}

\end{document}